\documentclass[12pt,a4paper]{article}
\usepackage{amssymb}
\usepackage{amsmath}
\usepackage{latexsym}
\usepackage{cite}
\usepackage{physics}
\usepackage{float}
\usepackage{graphicx}
\begin{document}

\title{\vspace{-1cm}\bf Asymptotic behavior of solutions and spectrum\\ of states in the quantum scalar field theory\\ in the Schwarzschild spacetime}

\author{
Mikhail~N.~Smolyakov
\\
{\small{\em Skobeltsyn Institute of Nuclear Physics, Lomonosov Moscow
State University,
}}\\
{\small{\em Moscow 119991, Russia}}}

\date{}
\maketitle

\begin{abstract}
In this paper, the study of canonical quantization of a free real massive scalar field in the Schwarzschild spacetime is continued. The normalization constants for the eigenfunctions of the corresponding radial equation are calculated, providing the necessary coefficients for the doubly degenerate scatteringlike states that are used in the expansion of the quantum field. It is shown that one can pass to a new type of states such that the spectrum of states with energies larger than the mass of the field splits into two parts. The first part consists of states that resemble properly normalized plane waves far away from the black hole, so they just describe the theory for an observer located in that area. The second part consists of states that live relatively close to the horizon and whose wave functions decrease when one goes away from the black hole. The appearance of the second part of the spectrum, which follows from the initial degeneracy of the scatteringlike states, is a consequence of the topological structure of the Schwarzschild spacetime.
\end{abstract}

\section{Introduction}
The problem of quantization of fields in a curved background is widely discussed in the literature; see, for example, the well-known monograph \cite{BD} and references therein. Among the variety of backgrounds, one of the most interesting cases is a black hole, whereas the most famous effect related to black holes, for which a consistent quantum field theory is needed, is the Hawking effect. The simplest black hole solution is the Schwarzschild solution, for which the problem of field quantization is discussed in a huge number of papers, starting from the pioneering papers \cite{Boulware:1974dm,HH}. It is well known that, in order to describe the Hawking effect, it is necessary to take into account the areas both below and above the black hole horizon. The latter can be done by passing to the Kruskal–Szekeres coordinates \cite{Kruskal,Szekeres}, which describe the maximal analytic extension of the Schwarzschild spacetime. However, a mathematically rigorous approach to description of the Hawking effect \cite{Christensen:1977jc,Fulling:1977zs,Candelas:1980zt,Sciama:1981hr} relies on the knowledge of wave functions of states above the black hole horizon. Indeed, depending on which vacuum is supposed to be the physical one, i.e., the Boulware vacuum \cite{Boulware:1974dm}, the Hartle-Hawking vacuum \cite{HH}, or the Unruh vacuum \cite{Unruh:1976db,Unruh:1977ga}, the contribution of wave functions of states above the black hole horizon to the ``actual'' wave functions is different.

However, there is a controversy in the scientific literature concerning the properties of one-particle solutions of field equations for the simplest case of the scalar field. In particular, in the well-known paper \cite{Deruelle:1974zy}, it is stated that the spectrum of states of the corresponding radial equation for $E<M$ (here, $E$ is the energy of the state and $M$ is the mass of the field) is discrete (though each state has an infinite norm), whereas
in papers \cite{Zecca3,Barranco:2011eyw} it is shown that this part of the spectrum is continuous and the radial solutions can be expressed in terms of the Heun functions. In paper \cite{GN}, it is stated that from the quantum mechanical point of view the whole theory is ill behaved. So, in paper \cite{Egorov:2022hgg}, a detailed examination of solutions of the field equation in the case of a real massive scalar field was carried out and a rigorous procedure of canonical quantization in the area above the horizon was performed. It was explicitly demonstrated that the area below the horizon (i.e., the black hole itself) is indeed not necessary for obtaining a self-consistent quantum field theory (i.e., the theory in which the canonical commutation relations are satisfied exactly and the Hamiltonian has the standard form without pathologies) at least in the simplest case of the scalar field.\footnote{The quantum scalar field only outside the horizon of the Schwarzschild black hole has already been considered in the literature; see, for example, recent papers \cite{Akhmedov:2020ryq,Anempodistov:2020oki,Bazarov:2021rrb}.} This result is in agreement with the results presented in several recent papers by 't~Hooft \cite{tHooft1,tHooft2,tHooft3}, in which an attempt was made to solve some problems with the physical interpretation of the quantum theory in the presence of a black hole (taking into account the appearance of the second, so-called ``white hole'' and the well-known problem with locality due to the existence of the white hole in our Universe or even in a parallel world). Namely, in the approach proposed in \cite{tHooft1,tHooft2,tHooft3}, the interior regions of both holes do not play any role in the evolution and turn out to be mathematical artifacts that do not have a direct physical interpretation.\footnote{It should be noted that the approach has a drawback consisting in possible emergence of closed timelike curves \cite{tHooft3}.}

Usually, even when the quantum scalar field is considered only outside the horizon of the Schwarzschild black hole (again see, for example, \cite{Akhmedov:2020ryq,Anempodistov:2020oki,Bazarov:2021rrb}), the expansion of solutions of the field equation in spherical harmonics is used. In paper \cite{Egorov:2022hgg} the field is expanded in the scatteringlike states, which are close to slightly modified plane waves if we go far away from the black
hole, thus resembling wave functions of free particles in Minkowski spacetime.\footnote{The use of scattering states for examining scattering of scalar waves in the Schwarzschild metric can be found, for example, in \cite{Matzner,Andersson:1995vi}.} However, an unexpected result is that there exist {\em two} different states that look like slightly modified plane waves far away from the black hole. However, since the normalization constants for the radial solutions were not calculated in \cite{Egorov:2022hgg}, the coefficients in front of these modified plane waves were not calculated either. So, it was not clear what these modified plane waves correspond to.

In the present paper, I calculate the normalization constants for the eigenfunctions of the corresponding radial equation, which provide the necessary coefficients for the scatteringlike states. It turns out that, with explicit values of the coefficients, it becomes possible to combine two different scatteringlike states in such a way that there arise two different types of quantum states. Namely, the first type corresponds to states that resemble {\em properly normalized} plane waves far away from the black hole, so these states just describe the theory for an observer located in that area. The second type corresponds to states that live relatively close to the horizon and whose wave functions decrease when one goes away from the black hole.

This paper strongly relies on the results obtained in \cite{Egorov:2022hgg}. It is organized as follows. In Section~\ref{sectsetup}, the basic setup is described. In Section~\ref{sectnorm}, the normalization constants of the radial solutions are obtained. In Section~\ref{sectscatstates}, the coefficients for asymptotics of the scatteringlike states are obtained. In Section~\ref{sectnewstates}, new states are defined, which are even more useful for describing the quantum theory than the scatteringlike states that were introduced in \cite{Egorov:2022hgg}. In Section~\ref{sectqft}, the resulting quantum field theory is discussed. In Section~\ref{sectconclusion}, the main results obtained in the present paper are discussed. The Appendix contains auxiliary material.

\section{Setup}\label{sectsetup}
As in paper \cite{Egorov:2022hgg}, let us take a real massive scalar field $\phi(t,\vec x)$ in a curved background described by the Schwarzschild metric. First, we will consider the field at the classical level. Since the Schwarzschild metric is static, the equation of motion for the scalar field takes the form
\begin{equation}\label{scalareqm}
\sqrt{-g}\,g^{00}\ddot\phi+\partial_{i}\left(\sqrt{-g}\,g^{ij}\partial_{j}\phi\right)+M^{2}\sqrt{-g}\,\phi=0,
\end{equation}
where $\dot\phi=\partial_{0}\phi$. The scalar field $\phi(t,\vec x)$ can be expanded in solutions of the form
\begin{equation}\label{philm}
e^{\pm iE t}\phi_{lm}^{}(E,\vec x)=e^{\pm iE t}Y_{lm}(\theta,\varphi)f_{l}(E,r),
\end{equation}
where
\begin{equation}\label{Ylm}
Y_{lm}(\theta,\varphi)=\sqrt{\frac{2l+1}{4\pi}}\sqrt{\frac{(l-|m|)!}{(l+|m|)!}}\,P_{l}^{|m|}\left(\cos\theta\right)e^{im\varphi},\quad l=0,1,2, ... ,\quad m=0,\pm 1, \pm 2, ...
\end{equation}
are spherical harmonics in the convention of \cite{Korn-Korn}, leading to the radial equation
\begin{equation}\label{eqscalarrad}
E^{2}\frac{r}{r-r_{0}}f_{l}(E,r)-M^{2}f_{l}(E,r)+\frac{1}{r^{2}}\frac{d}{dr}\left(r(r-r_{0})\frac{df_{l}(E,r)}{dr}\right)
-\frac{l(l+1)}{r^{2}}f_{l}(E,r)=0
\end{equation}
for the functions $f_{l}(E,r)$. Without loss of generality, the functions $f_{l}(E,r)$ can be chosen to be real and one can set $E\ge 0$. Let us also restrict ourselves to the domain $r>r_{0}$, where $r_{0}$ is the Schwarzschild radius. Equation~\eqref{eqscalarrad} suggests the form of the orthogonality condition for $f_{l}(E,r)$, which is
\begin{equation}
\int\limits_{r_{0}}^{\infty}\frac{r^{3}}{r-r_{0}}f_{l}(E,r)f_{l}(E',r)\,dr=0\quad\textrm{for}\quad E\neq E',
\end{equation}
as well as the form of the norm
\begin{equation}\label{norm}
\int\limits_{r_{0}}^{\infty}\frac{r^{3}}{r-r_{0}}f_{l}^{2}(E,r)\,dr.
\end{equation}

\begin{figure}[ht]
\centering
\begin{minipage}{.49\textwidth}
\centering
\includegraphics[width=0.95\linewidth]{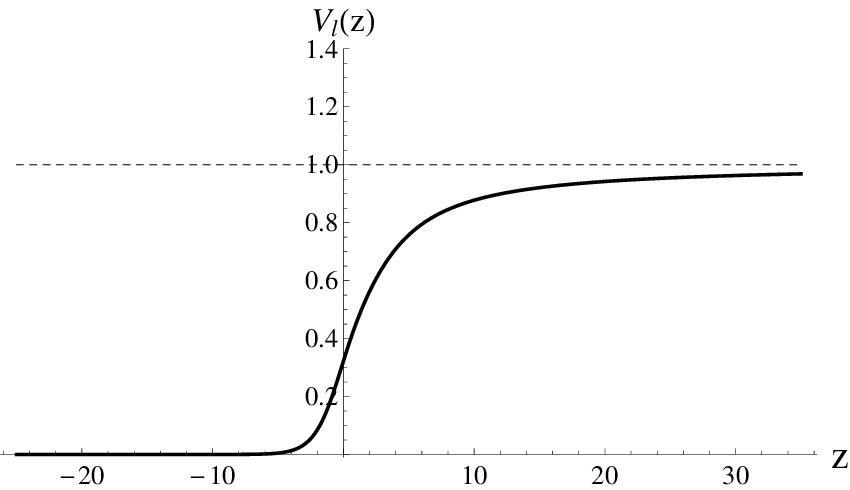}
\end{minipage}
\begin{minipage}{.49\textwidth}
\centering
\includegraphics[width=0.95\linewidth]{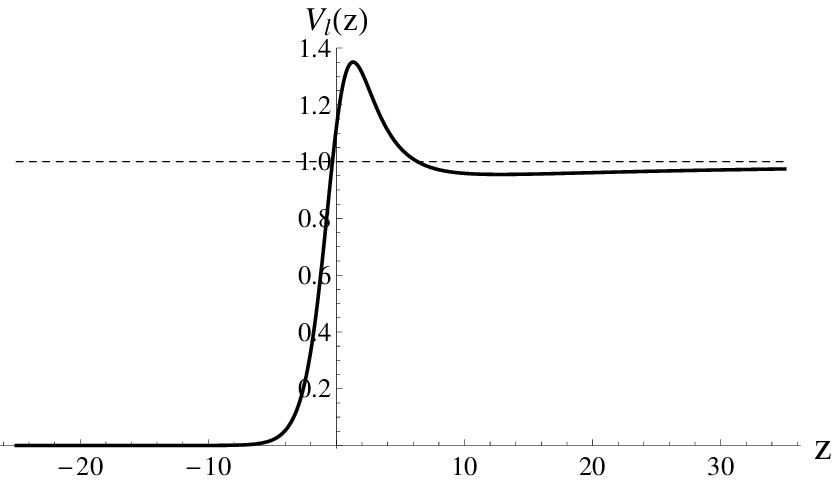}
\end{minipage}
\caption{$V_{l}(z)$ for $\mu=1$: $l=0$ (left plot) and $l=2$ (right plot). Dashed lines stand for $\mu^{2}$. The figure is taken from \cite{Egorov:2022hgg}.}\label{fig1}
\end{figure}

It is convenient to introduce the dimensionless variables and a new function:
\begin{equation}\label{substdimens}
\mu=Mr_{0},\qquad \epsilon=E r_{0},\qquad z=\frac{r}{r_{0}}+\ln\left(\frac{r}{r_{0}}-1\right), \qquad \psi_{l}(\epsilon,z)=rf_{l}(E,r).
\end{equation}
In these variables, Eq.~\eqref{eqscalarrad} can be expressed in the form of a one-dimensional Schr\"{o}dinger equation:
\begin{equation}\label{eqSchr}
-\frac{d^{2}\psi_{l}(\epsilon,z)}{dz^{2}}+V_{l}(z)\psi_{l}(\epsilon,z)=\epsilon^{2}\psi_{l}(\epsilon,z),
\end{equation}
where the potential has the form \cite{Barranco:2011eyw}
\begin{equation}\label{VSchr1}
V_{l}(z)=\frac{r(z)-r_{0}}{r(z)}\left(\mu^{2}+\frac{l(l+1)\,r_{0}^{2}}{r^{2}(z)}+\frac{r_{0}^{3}}{r^{3}(z)}\right)
\end{equation}
with $r(z)$ defined by \eqref{substdimens}. The potential $V_{l}(z)$ is such that $V_{l}(z)\to 0$ for $z\to-\infty$ and $V_{l}(z)\to\mu^{2}$ for $z\to\infty$. Since for $z\to\infty$ one gets $r(z)\approx r_{0}(z-\ln(z))$, at large $z$ the potential takes the form
\begin{equation}\label{potentialasymp}
V_{l}(z)\approx\mu^{2}\left(1-\frac{1}{z}\right).
\end{equation}
In Fig.~\ref{fig1}, some examples of $V_{l}(z)$ are presented.
One can see that it is a standard quantum mechanical problem, so the basic properties of the eigenfunctions (we are interested in physically relevant solutions, which are supposed to be bounded for $z\pm\infty$) are quite clear \cite{Barranco:2011eyw,Egorov:2022hgg}. Indeed, for $\epsilon<\mu$ and fixed $l$, there exists one solution parametrized by $q=\sqrt{\mu^{2}-\epsilon^{2}}$ with the asymptotics $\sim e^{-qz}$ at $z\to\infty$. For $\epsilon>\mu$ and fixed $l$ there exist two linearly independent solutions parametrized by $q=\sqrt{\epsilon^{2}-\mu^{2}}$. As was noted above, these solutions can be chosen to be real. Let us denote them by $\psi_{lp}(q,z)$, where $p=1,2$.

Applying transformations \eqref{substdimens} to the initial norm \eqref{norm}, we get
\begin{equation}
\int\limits_{r_{0}}^{\infty}\frac{r^{3}}{r-r_{0}}f_{l}^{2}(E,r)\,dr\to\int\limits_{-\infty}^{\infty}\psi_{l}^{2}(\epsilon,z)dz.
\end{equation}
This form of the norm also follows directly from Eq.~\eqref{eqSchr}. In what follows, the normalization condition
\begin{equation}\label{normq}
\int\limits_{-\infty}^{\infty}\psi_{lp}(q,z)\psi_{lp'}(q',z)dz=\delta_{pp'}\delta(q-q')
\end{equation}
will be used for solutions with $\epsilon>\mu$ instead of the condition
\begin{equation}\label{normepsilon}
\int\limits_{-\infty}^{\infty}\psi_{lp}(\epsilon,z)\psi_{lp'}(\epsilon',z)dz=\delta_{pp'}\delta(\epsilon-\epsilon')
\end{equation}
used in \cite{Egorov:2022hgg}.

\section{Normalization constants of the radial solutions}\label{sectnorm}
As was shown in \cite{Barranco:2011eyw}, for $\epsilon>\mu$ there exist two linearly independent solutions of Eq.~\eqref{eqSchr} with potential \eqref{VSchr1}. Let us find the asymptotics of these solutions. For $z\to\infty$, the equation takes the form
\begin{equation}\label{eqasympinfty}
-\frac{d^{2}\psi_{lp}(q,z)}{dz^{2}}+\mu^{2}\left(1-\frac{1}{z}\right)\psi_{lp}(q,z)=\epsilon^{2}\psi_{lp}(q,z),
\end{equation}
where only the leading terms in the potential $V_{l}(z)$ are retained. So, both solutions for $z\to\infty$ can be represented as
\begin{equation}\label{solz1}
\psi_{lp}(q,z)=C_{lp}^{+}(q)\sin\left(qz+\frac{\mu^{2}}{2q}\ln(z)+\kappa_{lp}(q)\right),
\end{equation}
where $\epsilon=\sqrt{q^{2}+\mu^{2}}$. Here, the unknown phases $\kappa_{lp}(q)$ can be defined in such a way that $C_{lp}^{+}(q)>0$. Note that the coefficients $C_{lp}^{+}(q)$ are also unknown at the moment. The term $\sim\ln(z)$ in \eqref{solz1} is due to the term $\frac{1}{z}$ in \eqref{eqasympinfty}. Analogously, for $z\to -\infty$ the equation takes the form
\begin{equation}
-\frac{d^{2}\psi_{lp}(q,z)}{dz^{2}}=\epsilon^{2}\psi_{lp}(q,z),
\end{equation}
where again only the leading terms are retained. Both solutions for $z\to-\infty$ can be represented as
\begin{equation}\label{solz2}
\psi_{lp}(q,z)=C_{lp}^{-}(q)\sin\left(\epsilon z+\gamma_{lp}(q)\right).
\end{equation}
Here, the unknown phases $\gamma_{lp}(q)$ can be defined in such a way that $C_{lp}^{-}(q)>0$; the coefficients $C_{lp}^{-}(q)$ are also unknown. In paper \cite{Egorov:2022hgg}, an assertion was made that these solutions can be chosen so that the normalization constants $C_{lp}^{+}(q)$ and $C_{lp}^{-}(q)$ do not depend on $l$. In such a case, one can write $C_{p}^{+}(q)$ and $C_{p}^{-}(q)$ for all $l$. However, explicit values of $C_{p}^{+}(q)$ and $C_{p}^{-}(q)$ were not calculated in \cite{Egorov:2022hgg}. Below, it will be shown that an even more stringent constraint can be imposed on $C_{lp}^{+}(q)$ and $C_{lp}^{-}(q)$ and explicit values of these constants will be obtained for solutions satisfying the constraint.

Let us figure out how the values of the constants $C_{lp}^{+}(q)$ and $C_{lp}^{-}(q)$ contribute to the normalization conditions \eqref{normq}. It is well known that, since the normalization integrals for eigenfunctions in such quantum mechanical systems diverge, the normalization constants are determined by the behavior of eigenfunctions in the asymptotic regions; see \cite{LL-QM}. So, below, we will use the trick that was used in \S21 of \cite{LL-QM} for calculating normalization constants in a similar case.

First, let us rewrite the integral in the lhs of \eqref{normq} as
\begin{align}\nonumber
&\int\limits_{-\infty}^{\infty}\psi_{lp}(q,z)\psi_{lp'}(q',z)dz\approx\int\limits_{-\infty}^{-L}C_{lp}^{-}(q)C_{lp'}^{-}(q')
\sin\left(\epsilon z+\gamma_{lp}(q)\right)\sin\left(\epsilon' z+\gamma_{lp'}(q')\right)dz\\\nonumber
+&\int\limits_{L}^{\infty}C_{lp}^{+}(q)C_{lp'}^{+}(q')
\sin\left(qz+\frac{\mu^{2}}{2q}\ln(z)+\kappa_{lp}(q)\right)\sin\left(q'z+\frac{\mu^{2}}{2q'}\ln(z)+\kappa_{lp'}(q')\right)dz\\\label{normderiv1}
+&\int\limits_{-L}^{L}\psi_{lp}(q,z)\psi_{lp'}(q',z)dz,
\end{align}
where $L$ is such that for $|z|>L$ asymptotic solutions \eqref{solz1} and \eqref{solz2} can be utilized with a sufficient accuracy. Since the third integral in the rhs of \eqref{normderiv1} is finite and can be neglected in comparison with an overall infinite value of the normalization integral, one can replace this finite integral in \eqref{normderiv1} by any other finite value, for example, by
\begin{align}\nonumber
&\int\limits_{-L}^{L}\psi_{lp}(q,z)\psi_{lp'}(q',z)dz\to\int\limits_{-L}^{0}C_{lp}^{-}(q)C_{lp'}^{-}(q')
\sin\left(\epsilon z+\gamma_{lp}(q)\right)\sin\left(\epsilon' z+\gamma_{lp'}(q')\right)dz\\\label{finiteint}+&
\int\limits_{0}^{L}C_{lp}^{+}(q)C_{lp'}^{+}(q')
\sin\left(qz+\frac{\mu^{2}}{2q}\ln(z)+\kappa_{lp}(q)\right)\sin\left(q'z+\frac{\mu^{2}}{2q'}\ln(z)+\kappa_{lp'}(q')\right)dz.
\end{align}
Then, for \eqref{normderiv1} we get
\begin{align}\nonumber
&\int\limits_{-\infty}^{\infty}\psi_{lp}(q,z)\psi_{lp'}(q',z)dz\\\nonumber
\approx&\int\limits_{-\infty}^{0}C_{lp}^{-}(q)C_{lp'}^{-}(q')
\sin\left(\epsilon z+\gamma_{lp}(q)\right)\sin\left(\epsilon' z+\gamma_{lp'}(q')\right)dz\\\label{normderiv2}
+&\int\limits_{0}^{\infty}C_{lp}^{+}(q)C_{lp'}^{+}(q')
\sin\left(qz+\frac{\mu^{2}}{2q}\ln(z)+\kappa_{lp}(q)\right)\sin\left(q'z+\frac{\mu^{2}}{2q'}\ln(z)+\kappa_{lp'}(q')\right)dz.
\end{align}

Now let us consider the second integral in the rhs of formula \eqref{normderiv2}. For $q\to q'$, in the leading order this integral can be rewritten as
\begin{align}\nonumber
&\int\limits_{0}^{\infty}C_{lp}^{+}(q)C_{lp'}^{+}(q')
\sin\left(qz+\frac{\mu^{2}}{2q}\ln(z)+\kappa_{lp}(q)\right)\sin\left(q'z+\frac{\mu^{2}}{2q'}\ln(z)+\kappa_{lp'}(q')\right)dz\\\nonumber
\approx&-\frac{1}{4}\int\limits_{0}^{\infty}C_{lp}^{+}(q)C_{lp'}^{+}(q)
\Biggl(e^{i\left(2qz+\frac{\mu^{2}}{q}\ln(z)+\kappa_{lp}(q)+\kappa_{lp'}(q)\right)}
+e^{-i\left(2qz+\frac{\mu^{2}}{q}\ln(z)+\kappa_{lp}(q)+\kappa_{lp'}(q)\right)}\\\label{normderiv3}
&-e^{i\left((q-q')z+\kappa_{lp}(q)-\kappa_{lp'}(q)\right)}-e^{i\left((q'-q)z+\kappa_{lp'}(q)-\kappa_{lp}(q)\right)}\Biggr)dz.
\end{align}
It is clear that the first two terms in the brackets in the rhs of formula \eqref{normderiv3} are purely oscillating and cannot contribute to the normalization integral. As for the last two terms, one easily gets
\begin{align}\nonumber
&\frac{1}{4}\int\limits_{0}^{\infty}C_{lp}^{+}(q)C_{lp'}^{+}(q)
\Biggl(e^{i\left((q-q')z+\kappa_{lp}(q)-\kappa_{lp'}(q)\right)}+e^{i\left((q'-q)z+\kappa_{lp'}(q)-\kappa_{lp}(q)\right)}\Biggr)dz\\\nonumber
=&\frac{1}{4}\int\limits_{-\infty}^{\infty}C_{lp}^{+}(q)C_{lp'}^{+}(q)
e^{i(q-q')z}\cos\left(\kappa_{lp}(q)-\kappa_{lp'}(q)\right)dz\\\label{normderiv4}
-&\frac{1}{2}\int\limits_{0}^{\infty}C_{lp}^{+}(q)C_{lp'}^{+}(q)\sin\left((q'-q)z\right)\sin\left(\kappa_{lp}(q)-\kappa_{lp'}(q)\right)dz.
\end{align}
We are interested in the limit $q\to q'$, in which $\sin\left((q'-q)z\right)\to 0$. So, the second integral in the rhs of \eqref{normderiv4} vanishes. As for the first integral in the rhs of \eqref{normderiv4}, one gets
\begin{align}\nonumber
&\frac{1}{4}\int\limits_{-\infty}^{\infty}C_{lp}^{+}(q)C_{lp'}^{+}(q)
e^{i(q-q')z}\cos\left(\kappa_{lp}(q)-\kappa_{lp'}(q)\right)dz\\
=&\frac{\pi}{2}\,C_{lp}^{+}(q)C_{lp'}^{+}(q)\cos\left(\kappa_{lp}(q)-\kappa_{lp'}(q)\right)\delta(q-q'),
\end{align}
leading to
\begin{align}\nonumber
&\int\limits_{0}^{\infty}C_{lp}^{+}(q)C_{lp'}^{+}(q')
\sin\left(qz+\frac{\mu^{2}}{2q}\ln(z)+\kappa_{lp}(q)\right)\sin\left(q'z+\frac{\mu^{2}}{2q'}\ln(z)+\kappa_{lp'}(q')\right)dz\\\label{normderiv5}
=&\frac{\pi}{2}\,C_{lp}^{+}(q)C_{lp'}^{+}(q)\cos\left(\kappa_{lp}(q)-\kappa_{lp'}(q)\right)\delta(q-q').
\end{align}

A fully analogous procedure can be performed for the first integral in the rhs of formula \eqref{normderiv2}, resulting in
\begin{align}\nonumber
&\int\limits_{-\infty}^{0}C_{lp}^{-}(q)C_{lp'}^{-}(q')
\sin\left(\epsilon z+\gamma_{lp}(q)\right)\sin\left(\epsilon' z+\gamma_{lp'}(q')\right)dz\\\nonumber&=\frac{\pi}{2}\,C_{lp}^{-}(q)C_{lp'}^{-}(q)\cos\left(\gamma_{lp}(q)-\gamma_{lp'}(q)\right)\delta(\epsilon-\epsilon')
\\\label{normderiv6}&=\frac{\pi}{2}\,C_{lp}^{-}(q)C_{lp'}^{-}(q)\cos\left(\gamma_{lp}(q)-\gamma_{lp'}(q)\right)\frac{\epsilon}{q}\,\delta(q-q').
\end{align}
Combining \eqref{normderiv5} and \eqref{normderiv6}, and taking into account \eqref{normq} and \eqref{normderiv2}, one can get
\begin{equation}
\frac{\pi}{2}\left(\frac{\epsilon}{q}\,C_{lp}^{-}(q)C_{lp'}^{-}(q)\cos\left(\gamma_{lp}(q)-\gamma_{lp'}(q)\right)
+C_{lp}^{+}(q)C_{lp'}^{+}(q)\cos\left(\kappa_{lp}(q)-\kappa_{lp'}(q)\right)\right)=\delta_{pp'}.
\end{equation}
The latter means that
\begin{equation}\label{normsingle}
\frac{\epsilon}{q}\,\left(C_{lp}^{-}(q)\right)^{2}+\left(C_{lp}^{+}(q)\right)^{2}=\frac{2}{\pi}
\end{equation}
for $p=p'$ and
\begin{equation}\label{orthocond}
\frac{\epsilon}{q}\,C_{l1}^{-}(q)C_{l2}^{-}(q)\cos\left(\gamma_{l1}(q)-\gamma_{l2}(q)\right)
+C_{l1}^{+}(q)C_{l2}^{+}(q)\cos\left(\kappa_{l1}(q)-\kappa_{l2}(q)\right)=0
\end{equation}
for $p\neq p'$.

There may arise a question concerning orthogonality condition \eqref{orthocond}. Indeed, a finite value of the integral in the lhs of \eqref{finiteint} can be neglected in comparison with the infinite value of the normalization integral, so it can be replaced by a different finite integral. However, one can think that this cannot be done in the case of two different eigenfunctions, because formally the finite integral cannot be neglected in comparison with the zero value of the orthogonality integral for these different eigenfunctions. The point is that, in the case of two linearly dependent solutions corresponding to the same eigenvalue, the orthogonality integral is also infinite, not finite. The latter can be easily checked by considering the standard sets of eigenfunctions with a continuous spectrum, for example, the one of the Fourier transform. Thus, the replacement in \eqref{finiteint} does not affect the derivation of the orthogonality condition \eqref{orthocond}, which, as the normalization condition \eqref{normsingle}, includes only the parameters of eigenfunctions at $z\to\pm\infty$. So, if condition \eqref{orthocond} is not fulfilled, the corresponding solutions are not orthogonal.

Recall that for two linearly independent solutions $\psi_{l1}(q,z)$ and $\psi_{l2}(q,z)$ the Wronskian is a constant:
\begin{equation}
\frac{d\psi_{l1}(q,z)}{dz}\,\psi_{l2}(q,z)-\frac{d\psi_{l2}(q,z)}{dz}\,\psi_{l1}(q,z)=\textrm{const},
\end{equation}
which means that
\begin{align}\nonumber
&\lim\limits_{z\to-\infty}\left(\frac{d\psi_{l1}(q,z)}{dz}\,\psi_{l2}(q,z)-\frac{d\psi_{l2}(q,z)}{dz}\,\psi_{l1}(q,z)\right)\\\label{Wronsksol}
=&\lim\limits_{z\to\infty}\left(\frac{d\psi_{l1}(q,z)}{dz}\,\psi_{l2}(q,z)-\frac{d\psi_{l2}(q,z)}{dz}\,\psi_{l1}(q,z)\right).
\end{align}
Because of \eqref{solz1} and \eqref{solz2}, relation \eqref{Wronsksol} results in
\begin{equation}\label{Wronsksol2}
\frac{\epsilon}{q}\,C_{l1}^{-}(q)C_{l2}^{-}(q)\sin\left(\gamma_{l1}(q)-\gamma_{l2}(q)\right)
=C_{l1}^{+}(q)C_{l2}^{+}(q)\sin\left(\kappa_{l1}(q)-\kappa_{l2}(q)\right).
\end{equation}
The latter relation will be used below.

Suppose that we have two orthogonal solutions $\psi_{l1}(q,z)$ and $\psi_{l2}(q,z)$. Then, the solutions
\begin{align}\label{transform1}
&\hat\psi_{l1}(q,z)=\cos\alpha\,\psi_{l1}(q,z)+\sin\alpha\,\psi_{l2}(q,z),\\
&\hat\psi_{l2}(q,z)=-\sin\alpha\,\psi_{l1}(q,z)+\cos\alpha\,\psi_{l2}(q,z)
\end{align}
are also orthogonal. Let us consider $\hat\psi_{l1}(q,z)$. Repeating the steps presented above for obtaining normalization condition \eqref{normsingle}, one can show that the contribution of the interval $z\to-\infty$ to the normalization integral is
\begin{equation}\label{area-}
\frac{\pi}{2}\frac{\epsilon}{q}\cos^{2}\alpha\left(\left(C_{l1}^{-}(q)\right)^{2}+\left(C_{l2}^{-}(q)\right)^{2}\tan^{2}\alpha
+2\,C_{l1}^{-}(q)\,C_{l2}^{-}(q)\cos\left(\gamma_{l1}(q)-\gamma_{l2}(q)\right)\tan\alpha\right),
\end{equation}
whereas the contribution of the interval $z\to\infty$ is
\begin{equation}\label{area+}
\frac{\pi}{2}\cos^{2}\alpha\left(\left(C_{l1}^{+}(q)\right)^{2}+\left(C_{l2}^{+}(q)\right)^{2}\tan^{2}\alpha
+2\,C_{l1}^{+}(q)\,C_{l2}^{+}(q)\cos\left(\kappa_{l1}(q)-\kappa_{l2}(q)\right)\tan\alpha\right).
\end{equation}
Combining these two contributions, one gets
\begin{align}\nonumber
&\frac{\epsilon}{q}\cos^{2}\alpha\left(\left(C_{l1}^{-}(q)\right)^{2}+\left(C_{l2}^{-}(q)\right)^{2}\tan^{2}\alpha
+2\,C_{l1}^{-}(q)\,C_{l2}^{-}(q)\cos\left(\gamma_{l1}(q)-\gamma_{l2}(q)\right)\tan\alpha\right)\\\label{areaboth}
&+\cos^{2}\alpha\left(\left(C_{l1}^{+}(q)\right)^{2}+\left(C_{l2}^{+}(q)\right)^{2}\tan^{2}\alpha
+2\,C_{l1}^{+}(q)\,C_{l2}^{+}(q)\cos\left(\kappa_{l1}(q)-\kappa_{l2}(q)\right)\tan\alpha\right)=\frac{2}{\pi};
\end{align}
compare with its analog \eqref{normsingle}.

Now let us ask the question whether it is possible to find such a value of the angle $\alpha$ that contributions of the intervals $z\to-\infty$ (formula \eqref{area-}) and $z\to\infty$ (formula \eqref{area+}) to normalization condition \eqref{areaboth} are proportional to each other with the same proportionality coefficient for any $l$. Note that the parameters $C_{lp}^{+}(q)>0$ (recall that $C_{lp}^{-}(q)$ can be expressed through $C_{lp}^{+}(q)$ by means of \eqref{normsingle}), $\kappa_{lp}(q)$, and $\gamma_{lp}(q)$ can be arbitrary. Thus, let us consider the relation
\begin{align}\nonumber
&\frac{\epsilon}{q}\left(\left(C_{l1}^{-}(q)\right)^{2}+\left(C_{l2}^{-}(q)\right)^{2}\tan^{2}\alpha
+2\,C_{l1}^{-}(q)\,C_{l2}^{-}(q)\cos\left(\gamma_{l1}(q)-\gamma_{l2}(q)\right)\tan\alpha\right)\beta^{2}(q)\\
=&\left(C_{l1}^{+}(q)\right)^{2}+\left(C_{l2}^{+}(q)\right)^{2}\tan^{2}\alpha
+2\,C_{l1}^{+}(q)\,C_{l2}^{+}(q)\cos\left(\kappa_{l1}(q)-\kappa_{l2}(q)\right)\tan\alpha,
\end{align}
where $\beta(q)$ is the proportionality coefficient which does not depend on $l$. This relation is just a quadratic equation for $\tan\alpha$:
\begin{align}\nonumber
&\tan\alpha\left(2\,C_{l1}^{+}(q)\,C_{l2}^{+}(q)\cos\left(\kappa_{l1}(q)-\kappa_{l2}(q)\right)
-\beta^{2}(q)\frac{\epsilon}{q}2\,C_{l1}^{-}(q)\,C_{l2}^{-}(q)\cos\left(\gamma_{l1}(q)-\gamma_{l2}(q)\right)\right)\\
&+\left(C_{l1}^{+}(q)\right)^{2}-\beta^{2}(q)\frac{\epsilon}{q}\left(C_{l1}^{-}(q)\right)^{2}+\tan^{2}\alpha\left(\left(C_{l2}^{+}(q)\right)^{2}
-\beta^{2}(q)\frac{\epsilon}{q}\left(C_{l2}^{-}(q)\right)^{2}\right)=0.
\end{align}
Note that in the general case the angle $\alpha$ depends on $l$ and $q$ but the corresponding argument and subscript are skipped in order not to clutter up the formulas. The discriminant of this equation is
\begin{align}\nonumber
D&=4\left(C_{l1}^{+}(q)\,C_{l2}^{+}(q)\cos\left(\kappa_{l1}(q)-\kappa_{l2}(q)\right)
-\beta^{2}(q)\frac{\epsilon}{q}\,C_{l1}^{-}(q)\,C_{l2}^{-}(q)\cos\left(\gamma_{l1}(q)-\gamma_{l2}(q)\right)\right)^{2}\\\label{discriminant1}
&-4\left(\left(C_{l2}^{+}(q)\right)^{2}
-\beta^{2}(q)\frac{\epsilon}{q}\left(C_{l2}^{-}(q)\right)^{2}\right)
\left(\left(C_{l1}^{+}(q)\right)^{2}-\beta^{2}(q)\frac{\epsilon}{q}\left(C_{l1}^{-}(q)\right)^{2}\right).
\end{align}
It is not difficult to show that the terms in \eqref{discriminant1} can be rearranged in such a way that the discriminant takes the form
\begin{align}\nonumber
D&=4\Biggl(\beta^{2}(q)\frac{\epsilon}{q}\biggl(
C_{l1}^{+}(q)\,C_{l2}^{-}(q)\cos\left(\kappa_{l1}(q)-\kappa_{l2}(q)\right)
-C_{l2}^{+}(q)\,C_{l1}^{-}(q)\cos\left(\gamma_{l1}(q)-\gamma_{l2}(q)\right)
\biggr)^{2}\\\nonumber
&+\left(C_{l1}^{+}(q)\right)^{2}\left(\beta^{2}(q)\frac{\epsilon}{q}\left(C_{l2}^{-}(q)\right)^{2}
-\left(C_{l2}^{+}(q)\right)^{2}\right)\sin^{2}\left(\kappa_{l1}(q)-\kappa_{l2}(q)\right)\\\label{discriminant2}
&-\beta^{2}(q)\frac{\epsilon}{q}\left(C_{l1}^{-}(q)\right)^{2}\left(\beta^{2}(q)\frac{\epsilon}{q}\left(C_{l2}^{-}(q)\right)^{2}
-\left(C_{l2}^{+}(q)\right)^{2}\right)\sin^{2}\left(\gamma_{l1}(q)-\gamma_{l2}(q)\right)\Biggr).
\end{align}
With the help of
\begin{equation}
\sin^{2}\left(\gamma_{l1}(q)-\gamma_{l2}(q)\right)
=\frac{q^{2}\left(C_{l1}^{+}(q)\right)^{2}\left(C_{l2}^{+}(q)\right)^{2}}
{\epsilon^{2}\left(C_{l1}^{-}(q)\right)^{2}\left(C_{l2}^{-}(q)\right)^{2}}
\sin^{2}\left(\kappa_{l1}(q)-\kappa_{l2}(q)\right),
\end{equation}
which follows from \eqref{Wronsksol2}, discriminant \eqref{discriminant2} can be brought to the form
\begin{align}\nonumber
&D=4\beta^{2}(q)\frac{\epsilon}{q}\Biggl(\biggl(
C_{l1}^{+}(q)\,C_{l2}^{-}(q)\cos\left(\kappa_{l1}(q)-\kappa_{l2}(q)\right)
-C_{l2}^{+}(q)\,C_{l1}^{-}(q)\cos\left(\gamma_{l1}(q)-\gamma_{l2}(q)\right)
\biggr)^{2}\\\nonumber
&+\frac{\left(C_{l1}^{+}(q)\right)^{2}}{\left(C_{l2}^{-}(q)\right)^{2}}\sin^{2}\left(\kappa_{l1}(q)-\kappa_{l2}(q)\right)\\&\times
\biggl(\frac{1}{\beta^{2}(q)}\frac{q}{\epsilon}\left(C_{l2}^{+}(q)\right)^{2}-\left(C_{l2}^{-}(q)\right)^{2}\biggr)
\biggl(\beta^{2}(q)\frac{q}{\epsilon}\left(C_{l2}^{+}(q)\right)^{2}
-\left(C_{l2}^{-}(q)\right)^{2}\biggr)\Biggr).
\end{align}
One can easily see that, for $\beta^{2}(q)\equiv 1$,
\begin{align}\nonumber
&D=4\frac{\epsilon}{q}\Biggl(\biggl(
C_{l1}^{+}(q)\,C_{l2}^{-}(q)\cos\left(\kappa_{l1}(q)-\kappa_{l2}(q)\right)
-C_{l2}^{+}(q)\,C_{l1}^{-}(q)\cos\left(\gamma_{l1}(q)-\gamma_{l2}(q)\right)
\biggr)^{2}\\
&+\frac{\left(C_{l1}^{+}(q)\right)^{2}}{\left(C_{l2}^{-}(q)\right)^{2}}
\biggl(\frac{q}{\epsilon}\left(C_{l2}^{+}(q)\right)^{2}-\left(C_{l2}^{-}(q)\right)^{2}\biggr)^{2}
\sin^{2}\left(\kappa_{l1}(q)-\kappa_{l2}(q)\right)\Biggr)\ge 0
\end{align}
for any $C_{l1}^{+}(q)$ and $C_{l2}^{+}(q)$. It means that for any parameters of asymptotic solutions \eqref{solz1} and \eqref{solz2} there always exists such a value of angle $\alpha$ (recall that in the general case $\alpha$ depends on $l$ and $q$) that the proportionality coefficient $\beta(q)$ is the same for all $l$ and $q$ and is just equal to unity.

It is clear that the solution $\hat\psi_{l1}(q,z)$ can be represented in the form
\begin{align}\label{hatpsirepres1}
&\hat\psi_{l1}(q,z)=\hat C_{l1}^{+}(q)\sin\left(qz+\frac{\mu^{2}}{2q}\ln(z)+\hat\kappa_{l1}(q)\right)\quad\textrm{for}\quad z\to\infty,\\
&\hat\psi_{l1}(q,z)=\hat C_{l1}^{-}(q)\sin\left(\epsilon z+\hat\gamma_{l1}(q)\right)\quad\textrm{for}\quad z\to-\infty,
\end{align}
where the phases $\hat\kappa_{l1}(q)$ and $\hat\gamma_{l1}(q)$ can be defined in such a way that $\hat C_{l1}^{+}(q)>0$ and $\hat C_{l1}^{-}(q)>0$. The fact that $\beta^{2}(q)\equiv 1$ for this solution implies that
\begin{equation}\label{p1cond1}
\frac{\epsilon}{q}\,\left(\hat C_{l1}^{-}(q)\right)^{2}=\left(\hat C_{l1}^{+}(q)\right)^{2}.
\end{equation}
On the other hand, the normalization condition implies
\begin{equation}\label{p1cond2}
\frac{\epsilon}{q}\,\left(\hat C_{l1}^{-}(q)\right)^{2}+\left(\hat C_{l1}^{+}(q)\right)^{2}=\frac{2}{\pi};
\end{equation}
see \eqref{normsingle}. Combining \eqref{p1cond1} and \eqref{p1cond2}, we finally get
\begin{equation}\label{C1result}
\hat C_{l1}^{+}(q)=\frac{1}{\sqrt{\pi}},\qquad \hat C_{l1}^{-}(q)=\sqrt{\frac{q}{\pi\epsilon}}.
\end{equation}

The orthogonal solution $\hat\psi_{l2}(q,z)$ can be also represented in the analogous form
\begin{align}
&\hat\psi_{l2}(q,z)=\hat C_{l2}^{+}(q)\sin\left(qz+\frac{\mu^{2}}{2q}\ln(z)+\hat\kappa_{l2}(q)\right)\quad\textrm{for}\quad z\to\infty,\\
&\hat\psi_{l2}(q,z)=\hat C_{l2}^{-}(q)\sin\left(\epsilon z+\hat\gamma_{l2}(q)\right)\quad\textrm{for}\quad z\to-\infty,
\end{align}
where $\hat C_{l2}^{+}(q)>0$ and $\hat C_{l2}^{-}(q)>0$. With \eqref{C1result}, it follows from the orthogonality condition \eqref{orthocond} and condition \eqref{Wronsksol2} that
\begin{align}
&\sqrt{\frac{\epsilon}{q}}\,\hat C_{l2}^{-}(q)\cos\left(\hat\gamma_{l1}(q)-\hat\gamma_{l2}(q)\right)=
-\hat C_{l2}^{+}(q)\cos\left(\hat\kappa_{l1}(q)-\hat\kappa_{l2}(q)\right),\\
&\sqrt{\frac{\epsilon}{q}}\,\hat C_{l2}^{-}(q)\sin\left(\hat\gamma_{l1}(q)-\hat\gamma_{l2}(q)\right)
=\hat C_{l2}^{+}(q)\sin\left(\hat\kappa_{l1}(q)-\hat\kappa_{l2}(q)\right),
\end{align}
leading to
\begin{equation}\label{p2cond1}
\frac{\epsilon}{q}\,\left(\hat C_{l2}^{-}(q)\right)^{2}=\left(\hat C_{l2}^{+}(q)\right)^{2}.
\end{equation}
On the other hand, from the normalization condition (again, see \eqref{normsingle}) it follows that
\begin{equation}\label{p2cond2}
\frac{\epsilon}{q}\,\left(\hat C_{l2}^{-}(q)\right)^{2}+\left(\hat C_{l2}^{+}(q)\right)^{2}=\frac{2}{\pi}.
\end{equation}
Combining \eqref{p2cond1} and \eqref{p2cond2}, we finally get
\begin{equation}
\hat C_{l2}^{+}(q)=\frac{1}{\sqrt{\pi}},\qquad \hat C_{l2}^{-}(q)=\sqrt{\frac{q}{\pi\epsilon}}.
\end{equation}

Thus, for $\epsilon>\mu$ one can always choose the set of eigenfunctions of problem \eqref{eqSchr} with \eqref{VSchr1} in such a form that
\begin{equation}\label{Cvaluesfinal}
C_{lp}^{+}(q)=\frac{1}{\sqrt{\pi}},\qquad C_{lp}^{-}(q)=\sqrt{\frac{q}{\pi\sqrt{q^{2}+\mu^{2}}}}
\end{equation}
for any $l$ and $p$.\footnote{An alternative derivation of \eqref{Cvaluesfinal} can be found in the Appendix.} One can see that this choice is even more stringent than the one used in \cite{Egorov:2022hgg}: Here, the coefficients $C_{lp}^{\pm}(q)$ do not depend not only on $l$ but on $p$ as well. Moreover, the coefficients $C_{lp}^{+}(q)$, which are necessary for the subsequent analysis, do not depend on $q$.

\section{Scatteringlike states}\label{sectscatstates}
Now let us return to the Schwarzschild coordinates and, with the help of \eqref{substdimens}, obtain the explicit form of $f_{lp}(k,r)$ from solutions \eqref{solz1} and \eqref{solz2}. Taking into account \eqref{Cvaluesfinal}, for $r\to\infty$ the result is
\begin{equation}\label{sol-large-r}
f_{lp}(k,r)\approx\frac{1}{\sqrt{\pi}\,r}\sin\left(kr+\frac{(2k^{2}+M^{2})r_{0}}{2k}\ln(kr)-\frac{\pi l}{2}+\tilde\delta_{lp}(k)\right),
\end{equation}
where $k=\frac{q}{r_{0}}$ and $\tilde\delta_{lp}(k)=\hat\kappa_{lp}(kr_{0})-\frac{(2k^{2}+M^{2})r_{0}}{2k}\ln(kr_{0})+\frac{\pi l}{2}$ are phase shifts; whereas for $r\to r_{0}$ the result is
\begin{equation}\label{sol-rr0}
f_{lp}(k,r)\approx\sqrt{\frac{k}{\pi\sqrt{k^{2}+M^{2}}}}\frac{1}{r_{0}}\sin\left(\sqrt{k^{2}+M^{2}}\,r_{0}\ln(k(r-r_{0}))+\tilde\gamma_{lp}(k)\right),
\end{equation}
where $\tilde\gamma_{lp}(k)=\hat\gamma_{lp}(k)+\sqrt{k^{2}+M^{2}}\,r_{0}\left(1-\ln(kr_{0})\right)$.

Now let us turn to the scatteringlike states. Let us define these states as
\begin{equation}\label{scatstatesdec}
\phi_{p}(\vec k,\vec x)=\frac{1}{4\pi k}\sum\limits_{l=0}^{\infty}(2l+1)e^{i\left(\frac{\pi l}{2}+\tilde\delta_{lp}(k)\right)}P_{l}\left(\frac{\vec k\vec x}{kr}\right)f_{lp}\left(k,r\right),
\end{equation}
where $P_{l}(...)$ are the Legendre polynomials, $\tilde\delta_{lp}(k)$ are phase shifts defined by representation \eqref{sol-large-r}, $k=|\vec k|$, $r=|\vec x|$, and $\vec n=\frac{\vec x}{r}$. Formula \eqref{scatstatesdec} differs from the one used in \cite{Egorov:2022hgg} only in the absence of the factor $\frac{\sqrt{k}}{\left(k^2+M^{2}\right)^{1/4}}$ --- this factor is not necessary here, because normalization condition \eqref{normq} is used from the very beginning instead of condition \eqref{normepsilon} that was used in \cite{Egorov:2022hgg}.

Using the results of \cite{Egorov:2022hgg}, one can easily show that at large $r$
\begin{equation}\label{scatstates}
\phi_{p}(\vec k,\vec x)\approx\frac{1}{\sqrt{2}(2\pi)^{\frac{3}{2}}}\left(e^{i\left(\vec k\vec x-\frac{(2k^{2}+M^{2})r_{0}}{2k}\ln(kr)\right)}+A_{p}(\vec k,\vec n,r)\frac{e^{ikr}}{r}\right),\qquad p=1,2,
\end{equation}
where the functions $A_{p}(\vec k,\vec n,r)$ are defined as \cite{Egorov:2022hgg}
\begin{equation}\label{scatampl}
A_{p}(\vec k,\vec n,r)=\frac{1}{2ik}\sum\limits_{l=0}^{\infty}(2l+1)P_{l}\left(\frac{\vec k\vec x}{kr}\right)\left(
e^{i\left(2\tilde\delta_{lp}(k)+\frac{(2k^{2}+M^{2})r_{0}}{2k}\ln(kr)\right)}-e^{-i\frac{(2k^{2}+M^{2})r_{0}}{2k}\ln(kr)}\right).
\end{equation}
The functions $A_{p}(\vec k,\vec n,r)$ look similar to the standard scattering amplitudes, but they explicitly depend on $r$ (pay attention to the slowly varying terms with $\ln(kr)$), so formally they are not actual scattering amplitudes. The extra slowly varying phase $\sim\ln(kr)$ in the plane wave solution $e^{i\left(\vec k\vec x-\frac{(2k^{2}+M^{2})r_{0}}{2k}\ln(kr)\right)}$ reflects the influence of the long-range potential $\sim\frac{1}{r}$, which is similar to the case of the standard Coulomb potential in quantum mechanics \cite{LL-QM}.

Note that, unlike the result obtained in \cite{Egorov:2022hgg}, here one gets formula \eqref{scatstates} with the exact value of the overall factor (the coefficient $\frac{1}{\sqrt{2}(2\pi)^{\frac{3}{2}}}$ in \eqref{scatstates}). As we will see in the next section, knowing this coefficient turns out to be important for defining new, more useful states.

\section{Passing to the new states $\phi_{+}$ and $\phi_{-}$}\label{sectnewstates}
The value of the overall coefficient $\frac{1}{\sqrt{2}(2\pi)^{\frac{3}{2}}}$ in \eqref{scatstates} suggests the following combinations of the scatteringlike states:
\begin{align}\label{phi+def}
&\phi_{+}(\vec k,\vec x)=\frac{1}{\sqrt{2}}\left(\phi_{1}(\vec k,\vec x)+\phi_{2}(\vec k,\vec x)\right),\\\label{phi-def}
&\phi_{-}(\vec k,\vec x)=\frac{1}{\sqrt{2}}\left(\phi_{1}(\vec k,\vec x)-\phi_{2}(\vec k,\vec x)\right).
\end{align}
From the results of \cite{Egorov:2022hgg}, it follows that these states satisfy the following orthogonality conditions:
\begingroup
\allowdisplaybreaks
\begin{align}\label{orthscatt2}
&\int\limits_{r>r_{0}}\sqrt{-g}\,g^{00}\phi_{lm}^{*}(E,\vec x)\phi_{\pm}^{}(\vec k,\vec x)\,d^{3}x=0,\\
&\int\limits_{r>r_{0}}\sqrt{-g}\,g^{00}\phi_{+}^{*}(\vec k,\vec x)\phi_{-}^{}(\vec k',\vec x)\,d^{3}x=0,\\\label{orthphi+}
&\int\limits_{r>r_{0}}\sqrt{-g}\,g^{00}\phi_{+}^{*}(\vec k,\vec x)\phi_{+}^{}(\vec k',\vec x)\,d^{3}x=\delta^{(3)}(\vec k-\vec k'),\\
&\int\limits_{r>r_{0}}\sqrt{-g}\,g^{00}\phi_{-}^{*}(\vec k,\vec x)\phi_{-}^{}(\vec k',\vec x)\,d^{3}x=\delta^{(3)}(\vec k-\vec k'),
\end{align}
\endgroup
where $\phi_{lm}^{}(E,\vec x)$ are defined by \eqref{philm} with \eqref{Ylm} and describe the states with $E<M$; see \cite{Egorov:2022hgg} for details. Together with $\phi_{lm}^{}(E,\vec x)$, they also form a complete set of eigenfunctions; the corresponding completeness relation can be easily obtained from the one found in \cite{Egorov:2022hgg} and takes the form
\begin{align}\nonumber
&\sum\limits_{l=0}^{\infty}\sum\limits_{m=-l}^{l}\int\limits_{0}^{M}\phi_{lm}^{*}(E,\vec x)\phi_{lm}^{}(E,\vec y)\,dE\\\label{complete4}
&+\int\phi_{-}^{*}(\vec k,\vec x)\phi_{-}^{}(\vec k,\vec y)\,d^{3}k+\int\phi_{+}^{*}(\vec k,\vec x)\phi_{+}^{}(\vec k,\vec y)\,d^{3}k=\frac{\delta^{(3)}(\vec x-\vec y)}{\sqrt{-g}\,g^{00}}.
\end{align}
Thus, any localized wave packet (such that it vanishes at $r\to r_{0}$ and at $r\to\infty$) at a fixed point in time can be expanded in the eigenfunctions $\phi_{lm}(E,\vec x)$, $\phi_{-}^{}(\vec k,\vec x)$ and $\phi_{+}^{}(\vec k,\vec x)$.

For large $r$, the functions $\phi_{+}(\vec k,\vec x)$ and $\phi_{-}(\vec k,\vec x)$ have the form
\begin{align}
&\phi_{+}(\vec k,\vec x)\approx\frac{1}{(2\pi)^{\frac{3}{2}}}\,e^{i\left(\vec k\vec x-\frac{(2k^{2}+M^{2})r_{0}}{2k}\ln(kr)\right)}
+\frac{1}{2(2\pi)^{\frac{3}{2}}}\left(A_{1}(\vec k,\vec n,r)+A_{2}(\vec k,\vec n,r)\right)\frac{e^{ikr}}{r},\\
&\phi_{-}(\vec k,\vec x)\approx\frac{1}{2(2\pi)^{\frac{3}{2}}}\left(A_{1}(\vec k,\vec n,r)-A_{2}(\vec k,\vec n,r)\right)\frac{e^{ikr}}{r}.
\end{align}
In particular, for $r\to\infty$, one can write
\begin{align}\label{phi+infty}
&\phi_{+}(\vec k,\vec x)\approx\frac{1}{(2\pi)^{\frac{3}{2}}}\,e^{i\left(\vec k\vec x-\frac{(2k^{2}+M^{2})r_{0}}{2k}\ln(kr)\right)},\\
&\phi_{-}(\vec k,\vec x)\approx 0.
\end{align}
A remarkable feature of \eqref{phi+infty} is that, apart from the term $\sim\ln(kr)$, this formula resembles the properly normalized eigenfunctions in the case of Minkowski spacetime:
\begin{equation}
\phi(\vec k,\vec x)=\frac{1}{(2\pi)^{\frac{3}{2}}}\,e^{i\vec k\vec x}.
\end{equation}
This similarity is very logical. Indeed, far away from the black hole, the spacetime is almost flat, and we expect that there should exist such a set of eigenfunctions that it resembles the set of plane waves of Minkowski spacetime in that area. Taking into account the fact that the larger $r$ is, the slower the term $\sim\ln(kr)$ varies with $r$, in a finite area at $r\to\infty$ the functions $\phi_{+}(\vec k,\vec x)$ are just plane waves with some extra phase. These arguments also suggest that the functions $A_{1}(\vec k,\vec n,r)$ and $A_{2}(\vec k,\vec n,r)$ are not singular.

Let us discuss a little more the states $\phi_{+}(\vec k,\vec x)$ and $\phi_{-}(\vec k,\vec x)$. Although the calculations that will be presented below are not mathematically rigorous, they can reveal some possible properties of the states under consideration. To begin with, let us consider the integral
\begin{equation}\label{normdelta+0}
\int\limits_{r>r_{0}}\sqrt{-g}\,g^{00}\phi_{+}^{*}(\vec k,\vec x)\phi_{+}^{}(\vec k,\vec x)\,d^{3}x=\delta^{(3)}(0).
\end{equation}
This relation follows directly from the orthogonality condition \eqref{orthphi+}. Let us consider such $r_{1}$ that $\sqrt{-g}\approx 1$ and $g^{00}\approx 1$ for $r\ge r_{1}$ with a good accuracy. Also, in the leading order $\phi_{+}^{*}(\vec k,\vec x)\phi_{+}^{}(\vec k,\vec x)\approx\frac{1}{(2\pi)^{3}}$ for $r\ge r_{1}$. Thus, one gets
\begin{equation}\label{normdelta+1}
\int\limits_{r>r_{1}}\sqrt{-g}\,g^{00}\phi_{+}^{*}(\vec k,\vec x)\phi_{+}^{}(\vec k,\vec x)\,d^{3}x
\approx\frac{4\pi}{(2\pi)^{3}}\int\limits_{r_{1}}^{\infty}r^{2}dr.
\end{equation}
Now let us consider the case of Minkowski spacetime. One gets
\begin{equation}
\delta^{(3)}(0)=\int\phi^{*}(\vec k,\vec x)\phi^{}(\vec k,\vec x)\,d^{3}x=\frac{4\pi}{(2\pi)^{3}}\int\limits_{0}^{\infty}r^{2}dr
=\frac{4\pi}{(2\pi)^{3}}\left(\int\limits_{0}^{r_{1}}r^{2}dr+\int\limits_{r_{1}}^{\infty}r^{2}dr\right).
\end{equation}
It is clear that the first integral in the rhs of the latter relation is finite, so it can be neglected in comparison with the infinite value of $\delta^{(3)}(0)$. So, we can write
\begin{equation}\label{normdeltaMink1}
\frac{4\pi}{(2\pi)^{3}}\int\limits_{r_{1}}^{\infty}r^{2}dr=\delta^{(3)}(0).
\end{equation}
Combining \eqref{normdelta+1} and \eqref{normdeltaMink1}, we arrive at
\begin{equation}\label{normdelta+2}
\int\limits_{r>r_{1}}\sqrt{-g}\,g^{00}\phi_{+}^{*}(\vec k,\vec x)\phi_{+}^{}(\vec k,\vec x)\,d^{3}x
=\delta^{(3)}(0).
\end{equation}

Formula \eqref{normdelta+2} implies that the area $r_{0}<r<r_{1}$ in the Schwarzschild spacetime does not give a significant contribution to the normalization integral \eqref{normdelta+0}. On the other hand, the area $r_{0}<r<r_{1}$ is not similar to the ball of radius $r_{1}$ in Minkowski spacetime providing a finite contribution to the normalization integral; it has a different topology. Indeed, now let us consider the integral
\begin{equation}\label{normdelta-0}
\int\limits_{r>r_{0}}\sqrt{-g}\,g^{00}\phi_{-}^{*}(\vec k,\vec x)\phi_{-}^{}(\vec k,\vec x)\,d^{3}x=\delta^{(3)}(0).
\end{equation}
Since $\phi_{-}^{*}(\vec k,\vec x)\phi_{-}^{}(\vec k,\vec x)\sim\frac{1}{r^{2}}$ for $r\ge r_{1}$, one cannot get somewhat proportional to $\delta^{(3)}(0)$ by taking the integral
\begin{equation}\label{normdelta-1}
\int\limits_{r>r_{1}}\sqrt{-g}\,g^{00}\phi_{-}^{*}(\vec k,\vec x)\phi_{-}^{}(\vec k,\vec x)\,d^{3}x.
\end{equation}
Indeed, the degree of divergence of this integral turns out to be smaller than the one of \eqref{normdeltaMink1}. But it means that, according to \eqref{normdelta-0},
\begin{equation}\label{normdelta-2}
\int\limits_{r_{0}<r<r_{1}}\sqrt{-g}\,g^{00}\phi_{-}^{*}(\vec k,\vec x)\phi_{-}^{}(\vec k,\vec x)\,d^{3}x=\delta^{(3)}(0).
\end{equation}
Thus, the reasoning presented above suggests that the state $\phi_{-}^{}(\vec k,\vec x)$ lives relatively close to the horizon. Here, the term ``relatively close'' is used in the sense that the wave function of the state falls off for $r\to\infty$ (as $\sim\frac{1}{r}$), but its decrease is not so fast as the exponential one of the state $\phi_{lm}^{}(E,\vec x)$ (for which one can say that it lives in the vicinity of the horizon as it looks like in the Schwarzschild coordinates). On the other hand, this reasoning also suggests that the state $\phi_{+}^{}(\vec k,\vec x)$ lives far away from the horizon, whereas its wave function falls off somehow as $r\to r_{0}$. Unfortunately, the behavior of the functions $\phi_{\pm}^{}(\vec k,\vec x)$ at $r\to r_{0}$ in more detail is unknown at the moment, because the phases $\tilde\gamma_{lp}(k)$ in radial solutions \eqref{sol-rr0} are still unknown.

\section{Quantum theory}\label{sectqft}
Now we are ready to consider the scalar field at the quantum level. As was already mentioned in the Introduction, usually the quantum scalar field $\phi(t,\vec x)$ is expanded in spherical harmonics when the Schwarzschild spacetime is considered. In paper \cite{Egorov:2022hgg}, a different expansion was used, which is
\begin{align}\nonumber
\phi(t,\vec x)=\sum\limits_{l=0}^{\infty}\sum\limits_{m=-l}^{l}\int\limits_{0}^{M}\frac{dE}{\sqrt{2E}}\left(e^{-iEt}\phi_{lm}^{}(E,\vec x)a_{lm}^{}(E)+e^{iEt}\phi_{lm}^{*}(E,\vec x)a_{lm}^{\dagger}(E)\right)&\\\label{operatordec}
+\sum\limits_{p=1}^{2}\int\frac{d^{3}k}{\sqrt{2\sqrt{k^{2}+M^{2}}}}\left(e^{-i\sqrt{k^{2}+M^{2}}\,t}\phi_{p}^{}(\vec k,\vec x)a_{p}^{}(\vec k)+
e^{i\sqrt{k^{2}+M^{2}}\,t}\phi_{p}^{*}(\vec k,\vec x)a_{p}^{\dagger}(\vec k)\right)&,
\end{align}
where $\phi_{lm}^{}(E,\vec x)$ is defined by \eqref{philm} with \eqref{Ylm} and $\phi_{p}^{}(\vec
k,\vec x)$ is defined by \eqref{scatstatesdec}. In this expansion, the creation and annihilation operators satisfy the standard commutation relations
\begin{align}\label{CRa1}
&[a_{lm}^{}(E),a_{l'm'}^{\dagger}(E')]=\delta_{ll'}\delta_{mm'}\delta(E-E'),\\\label{CRa2}
&[a_{p}^{}(\vec k),a_{p'}^{\dagger}({\vec k}')]=\delta_{pp'}\delta^{(3)}(\vec k-\vec k'),
\end{align}
all other commutators being equal to zero. However, it is easy to see that with \eqref{phi+def} and \eqref{phi-def} expansion \eqref{operatordec} can be transformed into
\begin{align}\nonumber
\phi(t,\vec x)=\sum\limits_{l=0}^{\infty}\sum\limits_{m=-l}^{l}\int\limits_{0}^{M}\frac{dE}{\sqrt{2E}}\left(e^{-iEt}\phi_{lm}^{}(E,\vec x)a_{lm}^{}(E)+e^{iEt}\phi_{lm}^{*}(E,\vec x)a_{lm}^{\dagger}(E)\right)&\\\nonumber
+\int\frac{d^{3}k}{\sqrt{2\sqrt{k^{2}+M^{2}}}}\left(e^{-i\sqrt{k^{2}+M^{2}}\,t}\phi_{-}^{}(\vec k,\vec x)b^{}(\vec k)+
e^{i\sqrt{k^{2}+M^{2}}\,t}\phi_{-}^{*}(\vec k,\vec x)b^{\dagger}(\vec k)\right)&\\\label{operatordec2}
+\int\frac{d^{3}k}{\sqrt{2\sqrt{k^{2}+M^{2}}}}\left(e^{-i\sqrt{k^{2}+M^{2}}\,t}\phi_{+}^{}(\vec k,\vec x)a^{}(\vec k)+
e^{i\sqrt{k^{2}+M^{2}}\,t}\phi_{+}^{*}(\vec k,\vec x)a^{\dagger}(\vec k)\right)&,
\end{align}
where the new creation and annihilation operators are defined by means of the transformation
\begin{align}\label{opertrans1}
&a_{1}^{}(\vec k)=\frac{1}{\sqrt{2}}\left(a^{}(\vec k)+b^{}(\vec k)\right),\\\label{opertrans2}
&a_{2}^{}(\vec k)=\frac{1}{\sqrt{2}}\left(a^{}(\vec k)-b^{}(\vec k)\right).
\end{align}
One can easily check that for these new operators the standard commutation relations
\begin{align}
&[a^{}(\vec k),a^{\dagger}({\vec k}')]=\delta^{(3)}(\vec k-\vec k'),\\
&[b^{}(\vec k),b^{\dagger}({\vec k}')]=\delta^{(3)}(\vec k-\vec k')
\end{align}
hold, all other commutators also being equal to zero.

In the previous sections, though the theory was considered at the classical level, the term ``state'' was still used, since it was already assumed that the classical solutions discussed above would be related to the corresponding quantum states. In the present section we consider the quantum theory, so we should define the one-particle quantum states. Since the functions $\phi_{lm}(E,\vec x)$, $\phi_{-}^{}(\vec k,\vec x)$, and $\phi_{+}^{}(\vec k,\vec x)$ form a complete set of eigenfunctions, they can be used in determining the corresponding one-particle Hilbert space.\footnote{Since here the eigenfunctions have an infinite norm, strictly speaking one should consider the rigged Hilbert space \cite{rHs}.} For example, for $E>M$ the one-particle quantum states can be defined in the standard way as
\begin{align}
&\ket*{\vec k+}=\sqrt{2E}\,a^{\dagger}(\vec k)\ket{0},\\
&\ket*{\vec k-}=\sqrt{2E}\,b^{\dagger}(\vec k)\ket{0},
\end{align}
where $E=\sqrt{k^{2}+M^{2}}$. In such a case,
\begin{align}
&\bra{0}\phi(t,\vec x)\ket*{\vec k+}=e^{-iEt}\phi_{+}^{}(\vec k,\vec x),\\
&\bra{0}\phi(t,\vec x)\ket*{\vec k-}=e^{-iEt}\phi_{-}^{}(\vec k,\vec x)
\end{align}
are just the coordinate representations of the one-particle wave functions of the states $\ket*{\vec k+}$ and $\ket*{\vec k-}$, respectively. In particular, according to \eqref{phi+infty}, for $r\to\infty$ the wave function $\bra{0}\phi(t,\vec x)\ket*{\vec k+}$ behaves as a slightly modified plane wave, which is similar to the case of Minkowski spacetime.

Analogously, for $E<M$ the one-particle quantum states can be defined as
\begin{equation}
\ket{E,l,m}=\sqrt{2E}\,a_{lm}^{\dagger}(E)\ket{0}.
\end{equation}

Using the results presented in \cite{Egorov:2022hgg} for expansion \eqref{operatordec}, one can easily check that for expansion \eqref{operatordec2} the canonical commutation relations
\begin{equation}\label{CCR}
[\phi(t,\vec x),\pi(t,\vec y)]=i\delta^{(3)}(\vec x-\vec y),\qquad
[\phi(t,\vec x),\phi(t,\vec y)]=0,\qquad
[\pi(t,\vec x),\pi(t,\vec y)]=0,
\end{equation}
where the canonically conjugate momentum is
\begin{equation}\label{ccm}
\pi(t,\vec x)\equiv\frac{\partial\mathcal{L}}{\partial\dot\phi(t,\vec x)}=\sqrt{-g(\vec x)}\,g^{00}(\vec x)\dot\phi(t,\vec x),
\end{equation}
are exactly satisfied. The Hamiltonian of the system takes the form
\begin{equation}\label{Hamiltresult}
H=\sum\limits_{l=0}^{\infty}\sum\limits_{m=-l}^{l}\int\limits_{0}^{M}E\,a_{lm}^{\dagger}(E)a_{lm}^{}(E)\,dE
+\int\sqrt{k^{2}+M^{2}}
\left(b^{\dagger}(\vec k)b^{}(\vec k)+a^{\dagger}(\vec k)a^{}(\vec k)\right)d^{3}k,
\end{equation}
where the irrelevant $c$-number terms are dropped. Hamiltonian \eqref{Hamiltresult} can be easily obtained from the Hamiltonian derived in \cite{Egorov:2022hgg} with the help of \eqref{opertrans1} and \eqref{opertrans2}. It resembles the well-known Hamiltonian of the real scalar field in Minkowski spacetime:
\begin{equation}
H=\int\sqrt{k^{2}+M^{2}}\,a^{\dagger}(\vec k)a^{}(\vec k)\,d^{3}k.
\end{equation}
The difference between this Hamiltonian and the one in \eqref{Hamiltresult} is the existence of the states with $E<M$ that are localized near the horizon and the extra states with $E>M$ (the term with the operators $b^{}(\vec k)$ and $b^{\dagger}(\vec k)$).

\section{Discussion and conclusion}\label{sectconclusion}
In this paper, discussion of canonical quantization in the Schwarzschild spacetime is continued. In paper \cite{Egorov:2022hgg}, the quantum states for energies larger than the mass of the field were chosen such that they represent the scatteringlike states. A remarkable feature of the spectrum is that there exist two different scatteringlike states parametrized by the same asymptotic momentum $\vec k$. In the resulting theory, the canonical commutation relations are satisfied exactly and the Hamiltonian has the standard form. However, the coefficients in the asymptotics of the scatteringlike states were not calculated in \cite{Egorov:2022hgg}, so it was not clear what these scatteringlike states correspond to.

In the present paper, exact values of the corresponding coefficients for the scatteringlike states are calculated. The result suggests that for energies larger than the mass of the field it is more useful to pass to the different orthogonal quantum states $\phi_{+}(\vec k,\vec x)$ and $\phi_{-}(\vec k,\vec x)$ parametrized by the same asymptotic momentum $\vec k$, each being a linear combination of the two initial scatteringlike states. In the resulting theory, the canonical commutation relations are also satisfied exactly and the Hamiltonian also has the standard form. Since the theory is derived from the one obtained in \cite{Egorov:2022hgg}, the Schwarzschild black hole interior is not necessary for the resulting quantum field theory outside the black hole and does not affect it.

The states $\phi_{+}(\vec k,\vec x)$ and $\phi_{-}(\vec k,\vec x)$ have the following properties. The wave function of the state $\phi_{+}(\vec k,\vec x)$ is such that at $r\to\infty$ it looks like a slightly modified {\em properly normalized} plane wave. Meanwhile, some reasoning suggests that the wave function falls off for $r\to r_{0}$.\footnote{It should be mentioned once again that, though the reasoning presented in Section \ref{sectnewstates} (starting from formula \eqref{normdelta+0}) is not rigorous from the mathematical point of view, it can hint at possible properties of the states.} On the other hand, the wave function of the state $\phi_{-}(\vec k,\vec x)$ is such that it falls off as $\sim\frac{1}{r}$ at large $r$. Suppose that we have a localized wave packet that is located at large distance from the black hole. It is clear that contribution of the states $\phi_{-}(\vec k,\vec x)$ in the formation of the wave packet is negligible. Thus, if one considers scattering processes in some finite area far away from the black hole (of course, if the theory contains an interaction potential), only the states $\phi_{+}(\vec k,\vec x)$ would contribute to the corresponding processes, because the wave functions of $\phi_{-}(\vec k,\vec x)$ can be neglected in comparison with those of $\phi_{+}(\vec k,\vec x)$ at large $r$. Moreover, the states $\phi_{-}(\vec k,\vec x)$ would not show up as the virtual states, because contributions of the states $\phi_{-}(\vec k,\vec x)$ to the Green function
\begin{align}\nonumber
G(t,t',\vec x,\vec y)&=\sum\limits_{l=0}^{\infty}\sum\limits_{m=-l}^{l}\int\limits_{0}^{M}\int\limits_{0}^{M}\,dE\,d\tilde E\,
\frac{e^{-iE(t-t')}\phi_{lm}^{}(\tilde E,\vec x)\phi_{lm}^{*}(\tilde E,\vec y)}{2\pi\left({\tilde E}^{2}-E^{2}\right)}\\
&+\int\limits_{M}^{\infty}dE\int d^{3}k\,\frac{e^{-iE(t-t')}\left(\phi_{-}^{}(\vec k,\vec x)\phi_{-}^{*}(\vec k,\vec y)+\phi_{+}^{}(\vec k,\vec x)\phi_{+}^{*}(\vec k,\vec y)\right)}{2\pi\left(M^{2}+{\vec k}^{2}-E^{2}\right)}
\end{align}
can be also neglected at large $r$. Since the wave functions of the states $\phi_{+}(\vec k,\vec x)$ are just slightly modified plane waves, at large $r$ one gets a theory that is almost identical to the standard theory in Minkowski spacetime.

On the other hand, it looks as if the states $\phi_{-}(\vec k,\vec x)$ dominate close to the horizon, whereas contribution of the states $\phi_{+}(\vec k,\vec x)$ is suppressed in this area (of course, in this area the states with energies less than the mass of the field also show up, but they live much closer to the horizon and, thus, are not interesting for the present analysis). So, close to the horizon the theory is described by the states $\phi_{-}(\vec k,\vec x)$ (and, formally, also by the states $\phi_{lm}^{}(E,\vec x)$). One expects that there exists an intermediate zone, in which the wave functions of the states $\phi_{+}(\vec k,\vec x)$ and $\phi_{-}(\vec k,\vec x)$ are comparable, so both types of the states with energies larger than the mass of the field can contribute to the corresponding scattering processes.

Thus, we see that there exists a degeneracy of states parametrized by the same vector parameter $\vec k$. This degeneracy is a consequence of the degeneracy found in \cite{Egorov:2022hgg}, which, in turn, is a consequence of the topological structure $R^2\times S^2$ of the Schwarzschild spacetime (recall that the topological structure of Minkowski spacetime is $R^4$). Note that such a degeneracy of states is not expected for a very compact object, because such an object does not change the spacetime topology, but is expected for a traversable wormhole of the Morris-Thorne type \cite{Ellis:1973yv,Bronnikov:1973fh,Morris:1988cz} that connects two different universes \cite{Egorov:2022hgg}. Since the states $\phi_{-}(\vec k,\vec x)$ live relatively close to the horizon and are not seen directly by an observer located far away from the black hole, one can speculate that these states may constitute dark matter. This problem calls for a further analysis.

\subsection*{Acknowledgments}
The author is grateful to D.G.~Levkov and I.P.~Volobuev for valuable discussions, and to V.O.~Egorov, D.S.~Gorbunov, S.I.~Keizerov, E.R.~Rakhmetov, and G.I.~Rubtsov for useful comments. This study was conducted within the scientific program of the National Center for Physics and Mathematics, Section No.~5 ``Particle Physics and Cosmology'', stage 2023-2025.

\section*{Appendix: An alternative derivation of \eqref{Cvaluesfinal}}
Transformation \eqref{transform1} with \eqref{hatpsirepres1} implies that, for $r\to\infty$,
\begin{align}\nonumber
&\cos\alpha\,C_{l1}^{+}(q)\sin\left(qz+\frac{\mu^{2}}{2q}\ln(z)+\kappa_{l1}(q)\right)
+\sin\alpha\,C_{l2}^{+}(q)\sin\left(qz+\frac{\mu^{2}}{2q}\ln(z)+\kappa_{l2}(q)\right)\\&=\hat C_{l1}^{+}(q)\sin\left(qz+\frac{\mu^{2}}{2q}\ln(z)+\hat\kappa_{l1}(q)\right).
\end{align}
This relation can be rewritten as
\begin{align}\nonumber
&\Bigl(C_{l1}^{+}(q)\cos\alpha\cos\kappa_{l1}(q)
+C_{l2}^{+}(q)\sin\alpha\cos\kappa_{l2}(q)\Bigr)\sin\left(qz+\frac{\mu^{2}}{2q}\ln(z)\right)\\\nonumber +&\Bigl(C_{l1}^{+}(q)\cos\alpha\sin\kappa_{l1}(q)+C_{l2}^{+}(q)\sin\alpha\sin\kappa_{l2}(q)\Bigr)\cos\left(qz+\frac{\mu^{2}}{2q}\ln(z)\right)\\
=&\,\hat C_{l1}^{+}(q)\sin\left(qz+\frac{\mu^{2}}{2q}\ln(z)+\hat\kappa_{l1}(q)\right),
\end{align}
which leads to
\begin{align}\nonumber
&\Bigl(C_{l1}^{+}(q)\cos\alpha\cos\kappa_{l1}(q)
+C_{l2}^{+}(q)\sin\alpha\cos\kappa_{l2}(q)\Bigr)^{2}\\ +&\Bigl(C_{l1}^{+}(q)\cos\alpha\sin\kappa_{l1}(q)+C_{l2}^{+}(q)\sin\alpha\sin\kappa_{l2}(q)\Bigr)^{2}=\left(\hat C_{l1}^{+}(q)\right)^{2}.
\end{align}
The latter relation can be rewritten as
\begin{equation}
\left(C_{l1}^{+}(q)\right)^{2}\cos^{2}\alpha+\left(C_{l2}^{+}(q)\right)^{2}\sin^{2}\alpha+2\,C_{l1}^{+}(q)\,C_{l2}^{+}(q)
\sin\alpha\cos\alpha\cos\left(\kappa_{l1}(q)-\kappa_{l2}(q)\right)=\left(\hat C_{l1}^{+}(q)\right)^{2},
\end{equation}
which leads to the quadratic equation on $\tan\alpha$
\begin{align}\nonumber
&\left(\left(C_{l2}^{+}(q)\right)^{2}-\left(\hat C_{l1}^{+}(q)\right)^{2}\right)\tan^{2}\alpha+
2\,C_{l1}^{+}(q)\,C_{l2}^{+}(q)\cos\left(\kappa_{l1}(q)-\kappa_{l2}(q)\right)\tan\alpha\\\label{eqtanalphaapp}&
+\left(\left(C_{l1}^{+}(q)\right)^{2}-\left(\hat C_{l1}^{+}(q)\right)^{2}\right)=0.
\end{align}
The discriminant of this equation is
\begin{align}\nonumber
D&=4\left(C_{l1}^{+}(q)\right)^{2}\left(C_{l2}^{+}(q)\right)^{2}\cos^{2}\left(\kappa_{l1}(q)-\kappa_{l2}(q)\right)\\\nonumber
&-4\left(\left(C_{l2}^{+}(q)\right)^{2}-\left(\hat C_{l1}^{+}(q)\right)^{2}\right)
\left(\left(C_{l1}^{+}(q)\right)^{2}-\left(\hat C_{l1}^{+}(q)\right)^{2}\right)\\\nonumber
&=4\left(\hat C_{l1}^{+}(q)\right)^{2}\left(\left(C_{l1}^{+}(q)\right)^{2}+\left(C_{l2}^{+}(q)\right)^{2}\right)-4\left(\hat C_{l1}^{+}(q)\right)^{4}\\
&-4\left(C_{l1}^{+}(q)\right)^{2}\left(C_{l2}^{+}(q)\right)^{2}\sin^{2}\left(\kappa_{l1}(q)-\kappa_{l2}(q)\right).
\end{align}

Now let us consider relations \eqref{orthocond} and \eqref{Wronsksol2}. Using these relations, one can get
\begin{equation}
\frac{\epsilon^{2}}{q^{2}}\left(C_{l1}^{-}(q)\right)^{2}\left(C_{l2}^{-}(q)\right)^{2}=\left(C_{l1}^{+}(q)\right)^{2}\left(C_{l2}^{+}(q)\right)^{2}.
\end{equation}
Using \eqref{normsingle}, the latter relation can be rewritten as
\begin{equation}
\left(\frac{2}{\pi}-\left(C_{l1}^{+}(q)\right)^{2}\right)\left(\frac{2}{\pi}-\left(C_{l2}^{+}(q)\right)^{2}\right)=\left(C_{l1}^{+}(q)\right)^{2}\left(C_{l2}^{+}(q)\right)^{2},
\end{equation}
resulting in
\begin{equation}\label{C12sum}
\left(C_{l1}^{+}(q)\right)^{2}+\left(C_{l2}^{+}(q)\right)^{2}=\frac{2}{\pi}.
\end{equation}
With \eqref{C12sum}, for the discriminant one gets
\begin{equation}
D=\frac{8}{\pi}\left(\hat C_{l1}^{+}(q)\right)^{2}-4\left(\hat C_{l1}^{+}(q)\right)^{4}-4\left(C_{l1}^{+}(q)\right)^{2}\left(\frac{2}{\pi}-\left(C_{l1}^{+}(q)\right)^{2}\right)\sin^{2}\left(\kappa_{l1}(q)-\kappa_{l2}(q)\right).
\end{equation}
The maximal value of the term $\left(C_{l1}^{+}(q)\right)^{2}\left(\frac{2}{\pi}-\left(C_{l1}^{+}(q)\right)^{2}\right)$ is $\frac{1}{\pi^{2}}$, it is attained at $\left(C_{l1}^{+}(q)\right)^{2}=\frac{1}{\pi}$, so for the discriminant one can write
\begin{equation}
D\ge\frac{8}{\pi}\left(\hat C_{l1}^{+}(q)\right)^{2}-4\left(\hat C_{l1}^{+}(q)\right)^{4}-\frac{4}{\pi^{2}}=-4\left(\left(\hat C_{l1}^{+}(q)\right)^{2}-\frac{1}{\pi}\right)^{2}.
\end{equation}
Thus, $D\ge 0$ (which implies that there exists a solution of Eq.~\eqref{eqtanalphaapp} for any values of $C_{l1}^{+}(q)$, $\kappa_{l1}(q)$ and $\kappa_{l2}(q)$) for all $l$ only if $\hat C_{l1}^{+}(q)=\frac{1}{\sqrt{\pi}}$. Since relations \eqref{normsingle} and \eqref{C12sum} are valid for the coefficients $\hat C_{l1}^{+}(q)$ and $\hat C_{l2}^{+}(q)$ too, using these relations one can easily obtain \eqref{Cvaluesfinal}.

\end{document}